\documentclass[pra,showpacs,floatfix,amsmath,amsfonts,twocolumn]{revtex4}
\usepackage{graphics}
\usepackage{epsfig}
\usepackage{xspace}
\usepackage{color}

\newcommand{\bs}{\mathbf{s}}
\newcommand{\bPsi}{\mathbf{\Psi }}
\newcommand{\bPhi}{\mathbf{\Phi }}
\newcommand{\bpsi}{ \mbox{\boldmath$\psi$\unboldmath}  }
\newcommand{\bphi}{ \mbox{\boldmath$\phi$\unboldmath}  }

\newcommand{\bw}{{\bf w}}

\newcommand{\br}{\mathbf{r}}

\newcommand{\1}{|1 \rangle}

\newcommand*{\SOD}{SOD\@\xspace}

\begin{document}

\title{Bose-Einstein condensates with localized spin-orbit coupling: soliton complexes and spinor dynamics}

\author{Yaroslav V. Kartashov$^{1,2}$, Vladimir V. Konotop$^{3,4}$, and Dmitry A. Zezyulin$^{3}$}

\affiliation{$^{1}$ICFO-Institut de Ciencies Fotoniques, and Universitat Politecnica de Catalunya, 08860 Castelldefels (Barcelona), Spain
\\
$^{2}$Institute of Spectroscopy, Russian Academy of Sciences, Troitsk, Moscow Region, 
\\
$^{3}$Centro de F\'isica Te\'{o}rica e Computacional, Faculdade de Ci\^encias, Universidade de Lisboa, Avenida Professor Gama Pinto 2, Lisboa 1649-003, Portugal
\\
$^{4}$Departamento de
F\'isica, Faculdade de Ci\^encias, Universidade de Lisboa, Campo Grande, Edif\'icio C8,
Lisboa 1749-016, Portugal}

\date{\today}

\begin{abstract}

Spin-orbit (SO) coupling can be introduced in a Bose--Einstein condensate (BEC) as a gauge potential acting only in a localized spatial domain. Effect of such a SO ``defect'' can be understood by transforming the system  to the integrable vector model. The properties of the SO-BEC change drastically if the SO defect is accompanied by the Zeeman splitting. In such a non-integrable system, the SO defect qualitatively changes the character of soliton interactions and allows for  formation of stable \textit{nearly scalar} soliton complexes  with almost all atoms   concentrated in only one dark state. These solitons  exist only if the number of particles exceeds a  threshold value.
We also report on the possibility of transmission and reflection  of a soliton upon its scattering on the SO defect. Scattering strongly affects  the pseudo-spin polarization  and can induce  pseudo-spin precession. The scattering can also  result in almost complete atomic transfer between the dark states.

\end{abstract}
\maketitle

\section{Introduction}

Atomic gases in external fields represent a versatile tool for emulating phenomena  originally  predicted in other branches of physics, including  solid state physics~\cite{Lewenstein}, hydrodynamics~\cite{hydro}, theory of gravity~\cite{gravity}, optics~\cite{optics}, etc. Such systems allow for creation and control {\em in situ} of synthetic electric and magnetic fields, as well as potentials of practically any desirable shape. In this context, spin-orbit (SO) coupled Bose-Einstein condensates (BECs)~\cite{Nature,Galitski},  experimentally realized in~\cite{Spielman}, attract particular attention as they allow for studying phenomena related to the artificial vector gauge potentials~\cite{Delibard}.  Meantime, SO-BECs feature physical factors which are usually absent in the emulated systems. This is, in particular, the intrinsic nonlinearity of BEC, originating from inter-atomic interactions and supporting solitons in homogeneous BECs~\cite{Kevrekid,Wu} and in BECs with either Zeeman~\cite{KKA} or optical \cite{OL} lattices (both lattices are available experimentally~\cite{period_exper,OL_exper}).

More features of a SO-BEC can be explored due to flexibility of the SO-coupling. In particular, by using an external laser beam of a finite width one can implement a localized in space SO-coupling, i.e. a kind of {\em SO-coupling defect} (SOD). In this situation spinor components of the macroscopic wavefunction are coupled to the translational motion only in a localized spatial domain and are linearly decoupled outside it.
In the  absence of other external fields the effect of \SOD on stationary modes consists only in imprinting of spin texture and no scattering occurs when soliton interacts with \SOD, since the model remains integrable.
A remarkable fact, however, is that the situation changes dramatically  if  \SOD is created in a BEC  subjected to the Zeeman splitting. The system becomes non-integrable, the character of soliton interactions changes, and unique families of essentially nonlinear modes (multipole quasi-scalar complexes having no linear limit) appear. A soliton incident on the defect can be either transmitted or reflected (at weak or strong Zeeman fields, respectively), which is accompanied by precession of the pseudo-spin. These effects stemming from the interplay of the \SOD and Zeeman splitting constitute the subject of  the present paper.

The paper is organized as follows. The model is formulated in Sec.~\ref{sec:model}. In Sec.~\ref{sec:stationary} we study the bifurcation of stationary modes from the Manakov soliton solutions analytically (Sec.~\ref{sec:bifurc}), and families of the solutions and their stability numerically (Sec.~\ref{sec:numerics}). In Sec.~\ref{sec:scattering} we describe scattering of spinor solitons on the \SOD. Some technical details of calculations are given in the Appendixes.

\section{The model}
\label{sec:model}

We consider a cigar-shaped SO-BEC elongated in the $x$-direction and tightly bounded in the $(y,z)$-plane. The physical model for the coupling comes from the tripod scheme~\cite{Juzel}, having three ($j=1,2,3$) ground states $|j\rangle$ and one excited state $|0\rangle$  coupled by the laser beams $\Omega_{1,2}=2^{-1/2}\Omega_0 e^{-iky\mp i K(x)} \sin\theta $ and $\Omega_3=\Omega e^{ikz}\cos\theta$, where $k$ is the wavevector, $\theta$ and $\Omega_0$ are
constants
charactering the field amplitudes and phases, and $K(x)=\int_{-\infty}^{x}\kappa(\xi)d\xi $ is the phase modulation of the control beams.
The characteristic size of the condensate is considered much smaller than the diffraction length of the laser beams $\Omega_j$,
[even if $K(x)$ varies on the scale of a few microns], which allows one to neglect beam diffraction on the scale of the atomic cloud.

The linear part of the atomic Hamiltonian reads
\begin{eqnarray}
 H_{lin}=-\hbar\sum_{j=1}^3\Omega_j|0\rangle\langle j| + H.c.
\end{eqnarray}
and allows for the existence of dark states (see e.g.~\cite{Juzel})
\begin{eqnarray*}
\label{dark1}
|D_1\rangle=\frac{1}{\sqrt{2}} e^{ik(y+z)}\left(e^{iK(x)}|1\rangle -e^{-iK(x)}|2\rangle\right),
\\
\label{dark2}
|D_2\rangle=\frac{\cos\theta}{\sqrt{2}}  e^{ik(y+z)}\left(e^{iK(x)}|1\rangle+e^{-iK(x)}|2\rangle\right)
-\sin\theta |3\rangle.
\end{eqnarray*}
Now the $x$-component of the synthetic vector potential ${\bf A}=i\langle D_{m}(\br)|\nabla D_{n}(\br)\rangle$
(i.e. the Mead-Berry connection~\cite{MB,Juzel})
is computed as
$
A_x=-\kappa(x) \sigma_1
$
(hereafter $\sigma_{1,2,3}$ are the Pauli and $\sigma_0$ is the identity matrices and we use the dimensionless units defined by $m=\hbar=1$).
Accounting for a Zeeman field $\Omega$ and for attractive two-body interactions, we describe the quasi-1D SO-BEC by the spinor $\bPsi=(\Psi_1,\Psi_2)^T$ obeying the coupled Gross-Pitaevskii equations (GPEs)~\cite{Spielman,nonlinearity}
\begin{equation}
i\bPsi_t = \frac{1}{2}\left(\frac 1i {\frac{\partial }{\partial x }}-\kappa(x)\sigma_1\! \right)^2\!\!\bPsi
  +\frac{\Omega}{2} \sigma_3   \bPsi
 {-}(\bPsi^\dag\bPsi)\bPsi.
\label{GPE}
\end{equation}
This model is {\em exactly integrable} if either Zeeman splitting or SO coupling is taken into account along, but not both of them. If  $\kappa(x)\equiv 0$,   by the rotation $\bPhi =S_\Omega^{-1}(t)\bPsi$ with $S_\Omega(t)=e^{-i\Omega\sigma_3 t/2}$, Eq.~(\ref{GPE}) is reduced to the Manakov system (MS)~\cite{Manakov}
\begin{eqnarray}
\label{eq:Manakov}
i\bPhi_t =- \frac 12 \bPhi_{xx}-(\bPhi^\dag\bPhi)\bPhi,
\end{eqnarray}
 so that one-soliton solution of (\ref{GPE}) acquires the form $\bPsi =S_\Omega(t) \bPhi_M$ where
\begin{eqnarray}
\label{soliton}
\bPhi_{M}= \frac{\eta e^{ivx+i(\eta^2-v^2)t/2}}{\cosh[\eta(x-vt-x_0)]}
\left(\begin{array}{c}
e^{i\beta} \cos\alpha
\\
e^{-i\beta} \sin\alpha
\end{array}\right),
\end{eqnarray}
and $\eta$, $v$, $\alpha$, $\beta$, and $x_0$ are constants determining soliton parameters.
If $\Omega=0$, then the MS is obtained after the spatial rotation  $\bPsi =S_\kappa (x)\bPhi_M$ with
\begin{equation}
\label{spat_rot}
 S_\kappa(x)=\frac{1}{\sqrt{2}}\left(1-i\sigma_2\right)e^{i\sigma_3K}= \frac{1}{\sqrt{2}}\left(\!\begin{array}{cc} e^{iK(x)} & -e^{-iK(x)} \\ e^{iK(x)} & e^{-iK(x)} \end{array}\!\right)
\end{equation}

If $\kappa=$const and   $\Omega=0$  (or {\it vice versa} if $\Omega=$const and $\kappa=0$), then by the global rotation Eq.~(\ref{GPE}) can be rewritten in the form, in which stationary localized modes were thoroughly studied in optics~\cite{model_optics}. For the case of constant SO coupling, bright solitons were found for constant~\cite{Kevrekid} and periodic~\cite{KKA} Zeeman fields. Those solitons had two distinguishing features: in the limit of small number of atoms they bifurcated from the linear spectrum, and the populations of the dark states were comparable and even equal. These essentially vector solitons can form multi-hump complexes if repulsion between out-of-phase humps in one component is compensated by coupling with 
the second component.

The Zeeman field  applied simultaneously with SO coupling breaks gauge symmetry,  while spatial dependence of the SO-coupling, $\kappa(x)\ne  \textrm{const}$, breaks the translational symmetry. These broken symmetries lead to  much more complicated
stationary and dynamical properties of the condensate. \textit{Our first main result} is a new class of stationary modes having no counterparts in previously considered vector models and in scalar nonlinear Schr\"odinger (NLS) model (because of repulsion or attraction between neighboring solitons~\cite{Jianke}). These modes are
(i) multi-soliton complexes  {\em with no linear limit}, i.e. they require nonzero number of atoms; (ii)   \textit{nearly-scalar} which means that they are characterized by large population imbalance  between the spinor components
 (this is a counterintuitive situation as the linear coupling is supposed to act towards balancing the populations);
(iii) stable for properly chosen defect parameters.

Our \textit{second result} is the peculiar interaction of a moving vector soliton with the \SOD. We show almost complete transmission of a soliton through the defect at the Zeeman field below some critical value $\Omega<\Omega_{cr}$ and almost total reflection at $\Omega>\Omega_{cr}$. In both cases interaction of a soliton with the defect induces the pseudo-spin precession.

\section{The stationary problem}
\label{sec:stationary}

First we consider stationary modes: $\bPsi(x,t)=e^{-i\mu t}\bpsi(x)$, where $\mu$ is the chemical potential and $\bpsi(x)$ solves the stationary GPE
\begin{equation}
\mu \bpsi  = \frac{1}{2}\left(\frac 1i {\frac{\partial }{\partial x }}-\kappa(x)\sigma_1\! \right)^2\!\!\bpsi
+\frac{\Omega}{2} \sigma_3   \bpsi
{-}(\bpsi^\dag\bpsi)\bpsi.
\end{equation}
Even in the  absence of the Zeeman splitting ($\Omega=0$) the \SOD introduces inhomogeneous spinor texture because it couples two distinct spinor states at $x=\pm\infty$. This is seen from the local Stokes components $s_j=\bPsi^\dag\sigma_j\bPsi $, where $j=0,...,3$ and $\sigma_0$ is the identity matrix. Function $s_0(x,t)$ describes density of the condensate, and $s_1^2+s_2^2+s_3^2=s_0^2$. At $\Omega=0$ one obtains from (\ref{soliton}) that
\begin{eqnarray}
s_1=2s_0\cos(2\alpha),\quad s_3+is_2=-s_0\sin(2\alpha)e^{2i(K(x)+\beta)}
\end{eqnarray}
i.e., the pseudo-spin vector $\bs=(s_1,s_2,s_3)$ changes its orientation in the $(y,z)-$plane along the $x-$axis. The spin slope $\alpha$ and phase $\beta$ are arbitrary, so far.

\subsection{Nonlinear modes at small Zeeman field}
\label{sec:bifurc}

The situation changes when  $\Omega\neq 0$. To describe this case we perform the spatial rotation $\bpsi =S_\kappa (x)\bphi$ with $S_\kappa(x)$ defined in (\ref{spat_rot}) and obtain the system for the spinor $\bphi$:
\begin{eqnarray}
\mu\bphi = -\frac{1}{2}\frac{d^2\bphi}{d x^2}
-\frac{\Omega}{2} \sigma_1  e^{2i\sigma_3K(x)} \bphi
-(\bphi^\dag\bphi)\bphi.
\label{GPE_rotate}
\end{eqnarray}
When $\Omega=0$, a stationary  mode  of the latter equation localized at $x=0$, is obtained from the Manakov soliton (\ref{soliton}) at $v=x_0=0$: 
\begin{eqnarray}
\bphi_{0}=\frac{\eta}{\cosh(\eta x)}
\left(\begin{array}{c}
e^{i\beta}\cos\alpha
\\
e^{-i\beta}\sin\alpha
\end{array}\right), \quad  \eta = \sqrt{-\mu}.
\end{eqnarray}

Let us now consider the case with $|\Omega|\ll 1$. It is convenient to introduce the dimensionless variable $\xi=\eta x$, the spectral parameter $\nu=-\mu/\eta^2$, and to represent the wavefunction in the form (this representation as well as its convenience for the perturbation analysis is introduced and discussed in Ref.~\cite{Lakoba})
\begin{eqnarray}
\label{new_wave}
\bphi(x)=\eta e^{i\sigma_3\beta} S_\alpha  \bw(\xi),~~ S_\alpha=\left(\!\!
\begin{array}{cc}
\cos\alpha &-\sin\alpha
\\
\sin\alpha &\cos\alpha
\end{array}
\!\!\right).
\end{eqnarray}
It is straightforward to verify that for $\Omega \neq 0$ the vector $\bw$ solves the equation
\begin{eqnarray}
\label{w}
\frac{d^2\bw}{d\xi^2}+2(\bw^\dag\bw)\bw-\nu\bw=-\epsilon\hat{\omega}(\xi)\bw,
\end{eqnarray}
where $\epsilon=\Omega/\eta^2$,
\begin{eqnarray*}
\hat{\omega}(\xi)=\cos(2\alpha)\cos[Q(\xi)]\sigma_1+\sin[Q(\xi)]\sigma_2
\nonumber \\
+\sin(2\alpha)\cos[Q(\xi)]\sigma_3,
\end{eqnarray*}
and
\begin{equation}
\label{QK}
Q(\xi)=2[K(\xi/\eta)+\beta].
\end{equation}

Next, we set $\epsilon\ll 1$ and consider the expansion
\begin{subequations}
	\label{eq:small_amp}
\begin{eqnarray}
	\bw=\bw_0+\epsilon\left(\!
	\begin{array}{c} u_1(\xi) \\ v_1(\xi) \end{array}
	\!\right)+\epsilon^2\left(\!
	\begin{array}{c} u_2(\xi) \\ v_2(\xi) \end{array}
	\!\right)+\cdots,
	\\
	\nu=1+\epsilon\nu_1+\epsilon^2\nu_2+\cdots
\end{eqnarray}
\end{subequations}
where
\begin{eqnarray}
\label{w0}
\bw_0=\frac{1}{\cosh \xi}
\left(
\begin{array}{c} 1 \\ 0 \end{array}
\right).
\end{eqnarray}
The solvability conditions for the first order term $(u_1,v_1)^T$ of this expansion yield the constraints (see Appendix~\ref{app:bifurc} for the details)
\begin{subequations}
	\label{constratints}
	\begin{eqnarray}
	\label{cond2}
	\sin(2\alpha)\int_{-\infty}^{\infty} \frac{\cos[Q(\xi)]\sinh(\xi) d\xi}{\cosh^3(\xi)}=0,
	\\[2mm]
	\label{cond3}
	\int_{-\infty}^{\infty} \frac{\sin[Q(\xi)] d\xi}{\cosh^2(\xi)}=0,
	\\[2mm]
	\label{cond4}
	\cos(2\alpha)\int_{-\infty}^{\infty} \frac{\cos[Q(\xi)] d\xi}{\cosh^2(\xi)}=0.
	\end{eqnarray}
\end{subequations}

Now consider  the defect of the given parity, which in this section is understood as the parity of $\sin[Q(\xi)]$ and $\cos[Q(\xi)]$. Then for the existence of a family bifurcating from the stationary Manakov solution (at $\Omega=0$) constraint (\ref{cond2}) requires $\cos[Q(\xi)]=\cos[Q(-\xi)]$, while (\ref{cond3}) requires $\sin[Q(-\xi)]=-\sin[Q(\xi)]$. Finally, from (\ref{cond4}), where the integral is nonzero, we obtain that for the bifurcation of the family, the Manakov soliton must have $\alpha=\frac{\pi}{4}+\frac{\pi n}{2}$ ($n$ is an integer).

Summarizing the above results we conclude  that the families of solutions can bifurcate from the following Manakov solitons:
\begin{eqnarray}
\label{eq:branching}
\bphi_{j}=\frac{\eta}{\sqrt{2}\cosh(\eta x)}
\left(\begin{array}{c}
1
\\
(-1)^j
\end{array}\right),
\quad j=1,2
\end{eqnarray}
for an even defect $\kappa(x)=\kappa(-x)$, for which
\begin{equation}
Q(\xi) = 2[ K(x)-K(0)]=-2[K(-x)-K(0)],
\end{equation}
provided  that $\beta=-K(0)$. 

Notice that for an odd defect $\kappa(x)=-\kappa(-x)$ we have that  $K(x)-K(0)=K(-x)-K(0)$, and thus $Q(\xi)=Q(-\xi)$ and (\ref{cond3}) is generically not satisfied, i.e. there is no modes bifurcating from the Manakov soliton in the case of odd defect.

In the original field variables, (\ref{eq:branching}) means that  at small $\Omega\neq 0$ branching of a nonlinear mode is only possible from the vector solitons which at $\Omega=0$ read
\begin{equation}
\label{bifurc}
\bpsi_{1}= \frac{\sqrt{2}\eta}{\cosh(\eta x)}
\left(\!\!
\begin{array}{c}
\cos\{2[K(x)-K(0)]\}
\\
i\sin \{2[K(x)-K(0)]\}
\end{array}
\!\right)\!,
\,\,\,
\bpsi_{2}=\sigma_1\bpsi_{1}.
\end{equation}
Expressions (\ref{bifurc})   reveal some features characteristic to the nonlinear modes [see Fig.~\ref{fig:modes}(a,b)]: (i) there is $\pi/2$ phase shift between the components; (ii) density maximum at $x=0$ in one  component corresponds to the node of the another one; and (iii) decrease of the SO coupling ($K\to0$) results in a scalar soliton (all atoms are concentrated in one component).

 \begin{figure}
 \centering
 \includegraphics[width=0.9\columnwidth]{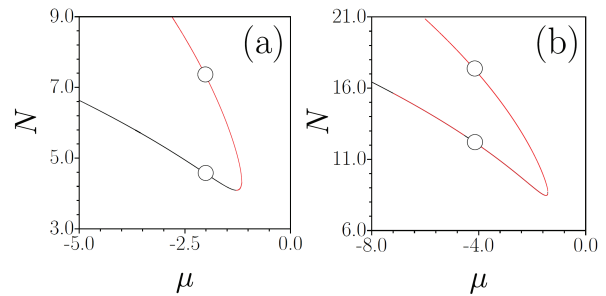}
 \caption{(Color online)    Families of monopole solitons for $w=1.6$ (a) and   dipole solitons for $w=1.5$ (b). In all cases $a=1$,  $\Omega=1$. Stable (unstable) families are shown in black (red). The circles correspond to solitons shown in Fig.~\ref{fig:modes}(a,c,e,g).}
 \label{fig:families}
 \end{figure}

\subsection{Numerical study of the nonlinear modes}
\label{sec:numerics}

To study the problem numerically, we focus on the  Gaussian \SOD $\kappa(x)= (2/\pi)^{-1/2}(a/w) e^{-x^2/(2w^2)}$, where $w$ is the width of the defect and $a$ determines its amplitude. 
When the SO coupling and Zeeman fields are fixed, soliton families can be characterized by the  dependence
of the number of atoms  $N=\int_{-\infty}^{\infty}s_0^2(x,t) dx$
{\it vs} $\mu$, see Fig.~\ref{fig:families}. We found that for a \SOD of finite width  all soliton families exist only if  the number of atoms exceeds a certain critical value $N_{cr}$.
This  can be understood from  Eq.~(\ref{GPE}) which includes
the \textit{expulsive} potential $\sim \kappa^2(x)$ induced by  the \SOD. Its influence can be compensated only by sufficiently strong attractive nonlinearity $\sim \bPsi^\dag\bPsi$ which requires  nonzero $N$.

\begin{figure*}[t]
	\includegraphics[width=\textwidth]{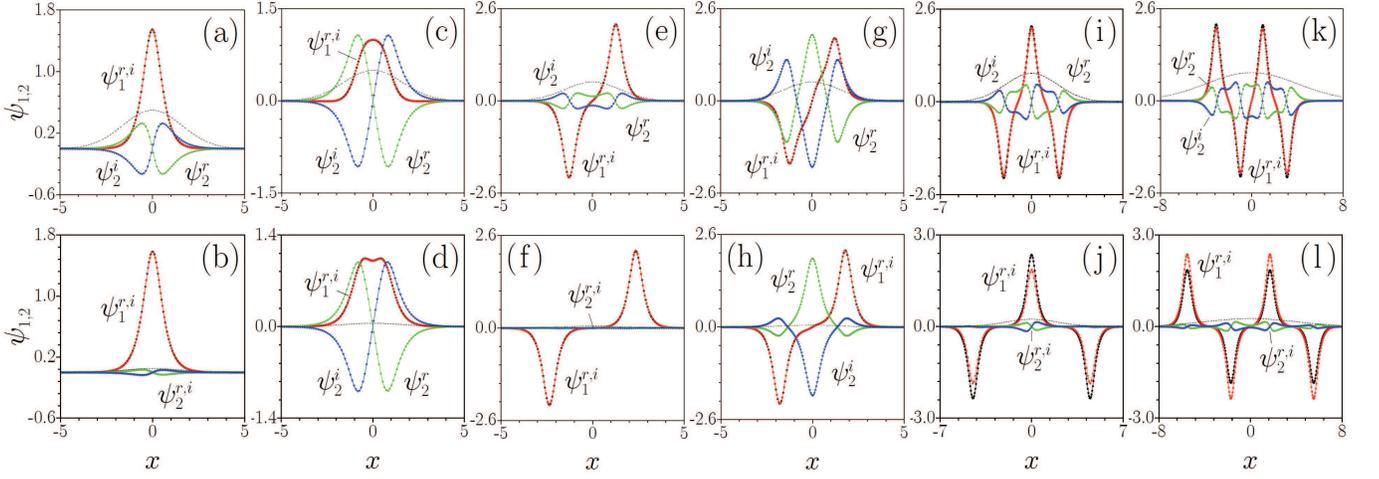}
	\caption{(Color online) Monopole modes from lower (a, b) and upper (c, d) branches at $\mu=-2$, $w=1.6$ shown in Fig.~\ref{fig:families}~(a). Panels (a, c) correspond to $a=1$, while panels (b, d) correspond to $a=0.1$. Dipole modes from the lower (e, f) and upper (g, h) branches   at $\mu=-2$, $w=1.5$ shown in Fig.~\ref{fig:families} (b). Panels (e, g) correspond to $a=1$, while panels (f, h) correspond to $a=0.1$.  Tripole solitons from the lower branch with $\mu=-4$, $w=2$, $a=2$ (i) and $a=0.6$ (j). Quadrupole solitons from the lower branch  with $\mu=-4$, $w=4$, $a=4$ (k) and $a=1.3$ (l). In all the cases $\Omega=1$.
Dashed lines show $\kappa(x)$ profiles.
	} \label{fig:modes}
\end{figure*}

\begin{figure}
 \centering
 \includegraphics[width=0.9\columnwidth]{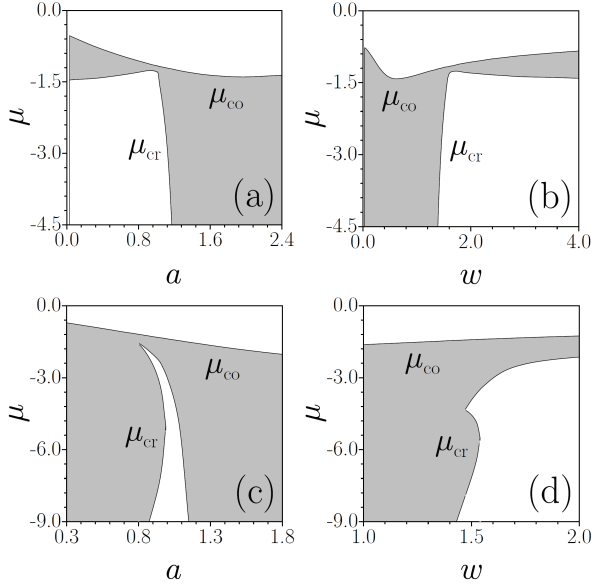}
 \caption{   Domains of stability (white) and instability (shaded) for the monopole (a, b) and dipole (c, d) solitons from the lower branches  in Fig.~\ref{fig:families}(a) and (b), respectively.
 In (a, c) the defect width  $w=1.6$, while in (b, d) the defect amplitude  $ a=1$. In all cases $\Omega= 1$.
} \label{fig:stability}
 \end{figure}

The found soliton families consist of the upper and lower branches joining at the cut-off value of the chemical potential $\mu_{co}$ (Figs.~\ref{fig:families} and \ref{fig:stability}). There is an infinite set of such families with progressively increasing complexity of soliton shapes.
  Solitons belonging to the lower branches incorporate one, two, or more (depending on the order of the family) out-of-phase humps in the first
 component and have rather complex structure of the second
  component [Figs.~\ref{fig:modes}(a,e,i,k)].
 For  solitons from the lower branch amplitude of the second component  can be  small compared to that of  the first component: say in Fig.~\ref{fig:modes}(e) the relation between the atomic density maxima in the states is $|\psi_2|^2/|\psi_1|^2\approx 0.010$. Therefore, these modes can be characterized as \emph{nearly scalar}.
  Solitons from the the upper branches resemble coupled monopole  and dipole modes [Fig.~\ref{fig:modes}(c)] for the first family, coupled dipole and tripole modes [Fig.~\ref{fig:modes}(g)] for the second family, etc.

Top and bottom rows of Fig.~\ref{fig:modes} illustrate the transformation of soliton profiles upon decrease of the SO coupling strength. In Figs.~\ref{fig:modes}(a,  b) one observes that the second component of  solitons from the lower branch nearly vanishes when $a\to 0$, i.e. one gets conventional (nearly-scalar) monopole soliton with almost all atoms concentrated in only one dark state. For $a=0$ this mode degenerates into the scalar soliton of the NLS  equation  and the threshold number of particles $N_{cr}$ vanishes. In contrast, the second component does not vanish for solitons from the upper branch even at $a\to 0$ [Fig.~\ref{fig:modes}(c,d)]; these solitons transform into fully vectorial solitons of the MS (after the rotation $S_\Omega(t)$ as explained above). The most unexpected result is shown in Figs.~\ref{fig:modes}(f,j,l) illustrating that decreasing strength of the \SOD results in gradual unfolding of the multi-hump solitons from the lower branch into sets of well-separated \textit{nearly scalar} solitons (the second component is hardly visible on the scale of Fig.~\ref{fig:modes}). This means that SO coupling {\em qualitatively} changes the character of soliton interactions: it  suppresses repulsion between out-of-phase humps (unavoidable in the scalar NLS equation~\cite{Jianke}), and allows for formation of nearly-scalar soliton complexes with arbitrary number of humps.

To understand qualitatively the effect of SO coupling which depends on the kinetic energy $H_{kin}=\frac 12 \int\bPsi^\dag p^2\bPsi dx$, where $p=-i\partial/\partial x$, let us consider nearly-scalar modes and address the simplest case of the constant coupling $\kappa=\kappa(0)$. Assuming that $|\Psi_2|\ll |\Psi_1|\ll 1$ (i.e. the weakly nonlinear limit) and small kinetic energy, for the stationary mode  we obtain $\psi_2\approx i\kappa \psi_{1,x}/(\mu+\Omega/2+\kappa^2/2)$. Taking into account that the phases of $\psi_{1,2}$ do not change with $x$ and differ by $\pi/2$, we obtain an estimate for the energy of interaction
\begin{equation}
H_{int}=-i\!\int\!\kappa\bpsi^\dag\sigma_1\bpsi_x dx\approx \frac{4\kappa^2}{2\mu+\Omega+\kappa^2}\int |\psi_{1x}|^2dx
\end{equation}
 For all considered modes $H_{int}<0$ and in the case if neighboring out-of-phase solitons the contribution from the term $|\psi_{1x}|$ to the integral increases  when two solitons approach each other. Therefore, the smaller is the separation between out-of-phase solitons, the smaller is $H_{int}<0$. Thus SO coupling diminishes the energy preventing decoupling of multihump solitons. The separation between humps in soliton complexes decreases with increase of the defect amplitude $a$ [Figs.~\ref{fig:modes}(e,i,k)]. We emphasize that the nearly-scalar states do not have analogs in previously considered vector models where repulsion between out-of-phase humps in one soliton component can only be compensated at expense of its coupling with nearly equally strong second component.

We also examined the linear  stability of the nonlinear modes (see Appendix~\ref{app:stabil}). We found that solitons from the upper branches are always unstable, but solitons from the lower branches can be stable in wide regions of their existence domain presented in Fig.~\ref{fig:stability}
for monopole  [Fig.~\ref{fig:stability}(a,b)] and dipole [Fig.~\ref{fig:stability}(c,d)] solitons. For the fundamental soliton  smaller defect amplitudes $a$ facilitate soliton stabilization, but the domains of stability may be rather complex   for multi-pole solitons  [Fig.~\ref{fig:stability}(c)]. Stability regions are also highlighted in Fig.~\ref{fig:families}.

\begin{figure}
\centering
\includegraphics[width=\columnwidth]{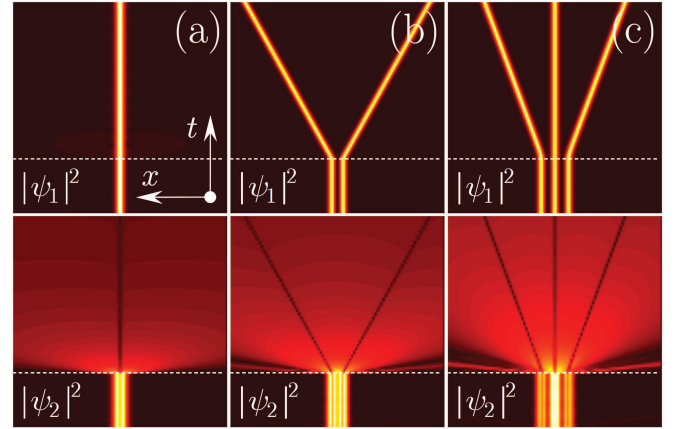}
\caption{(Color online) Splitting of stable modes
into solitons after switching off SO coupling at $t=50$ (dashed line).
The initial distributions correspond to the monopole   at $\mu=-2.5$, $a=1$, $w=1.6$ (a); dipole  at  $\mu=-4.2$, $a=1$, $w=1.5$ (b); and tripole  at $\mu=-4$, $a=1.5$, $w=3$ (c). In all cases $\Omega=1$ and the total evolution time is $t=200$. 
} \label{fig:split}
\end{figure}

Nearly-scalar multi-hump solitons exist also for a homogeneous SO coupling, which  in our case corresponds to $w\to\infty$ at $a/w=$const. However, linear stability analysis have shown that all such solitons [counterparts of states in Figs.~\ref{fig:modes}(e),(i),(k)] are unstable for all $\mu$ values as long as $\kappa=const$.
This analysis was conducted by solving the associated linear eigenvalue problem (see Supplementary Material). The structure of the spectrum, in particular, the presence of the eigenvalues with positive real part indicating on instability, are dictated only by the particular shape of spinor $\bPsi(x)$ and by the width of the $\kappa(x)$ function.
Therefore,   the finite width $w$ of the \SOD  is crucial, since it allows to stabilize the solitons. The fact that SO coupling is crucial for the formation of nearly-scalar multi-hump solitons is also illustrated in Fig.~\ref{fig:split} where abrupt switching off the SO coupling at $t=50$ results in the unfolding of stable multi-hump modes into a fan of diverging scalar NLS solitons.

\section{The scattering problem}
\label{sec:scattering}

At $x\to-\infty$ the \SOD vanishes and GPEs (\ref{GPE}) possess a solution $\bPsi=S_\Omega(t)\bPhi_M$. Now we consider interaction of this soliton moving with the initial velocity $v$ with the \SOD   located at $x=0$. To quantify scattering, we introduce the integral pseudo-spin components
$S_j(t)=\int_{-\infty}^{\infty}s_j(x,t)dx
$, $j=0,...,3$, which   before collision ($x\to-\infty$, designated by superscript ``$-$'') amount to
\begin{equation}
S_1^{-}+iS_2^{-}= 2\eta\sin(2\alpha)e^{i(\Omega t-\textcolor{black}{2\beta})}, ~~ S_3^{-} = 2\eta\cos(2\alpha)
\end{equation}
($S_0 = 2\eta$  is the conserved total number of atoms).
Notice that the integral components for the incident soliton satisfy the identity $[S_1^{-}]^2+[S_2^{-}]^2+[S_3^{-}]^2=S_0^2$, i.e.  one can say that the incident soliton features    pseudo-spin precession with   frequency $\Omega$.

\begin{figure}
	\centering
	\includegraphics[width=\columnwidth]{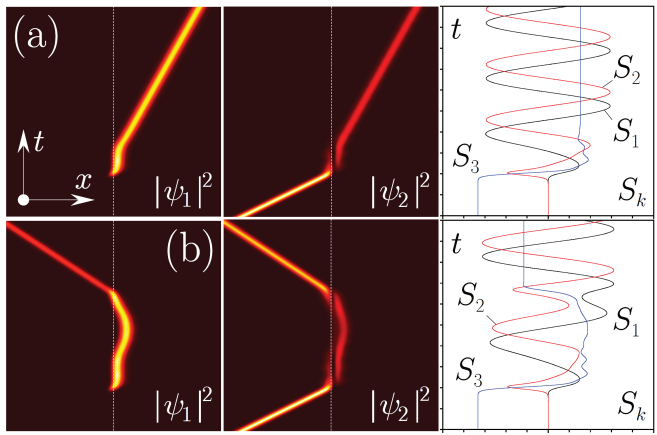}
	\caption{(Color online) Soliton interaction with the SOD in the presence of the Zeeman splitting $\Omega=0.099$ (a) and $\Omega=0.1$ (b). The parameters of the incident soliton and the SO defect are the same in both cases: $\eta=1$, $\alpha=\pi/2$, $v=0.4$, $\beta=0$, and $a=w=1$. The evolution is shown up to $t=240$ in the window $x\in[-24,24]$. The pseudo-spin components are shown in the window $S_k\in[-3,3]$.}
	\label{fig:scattering}
\end{figure}

At $\Omega=0$ no pseudo-spin precession occurs, and the soliton also does not undergo scattering,
%
because the rotation $S_\kappa^{-1}$
reduces Eq.~(\ref{GPE}) to the MS.
The situation changes in the presence of the Zeeman splitting ($\Omega>0$), as shown in Fig.~\ref{fig:scattering}, where
the initial solitons are chosen with  $\alpha=\pi/2$ so that all atoms populate only the second state.
In this case one observes \textit{either}
%
almost complete transmission through [Fig.~\ref{fig:scattering} (a)] or almost complete reflection by [Fig.~\ref{fig:scattering} (b)] the \SOD.
Although the existence of the transition region between transmission at smaller $\Omega$ and reflection at large $\Omega$ is expectable, a remarkable fact obtained numerically is a sharp  transition between the two scenarios  which occurs in extremely narrow domain of variations of the Zeeman field [$\Omega=0.099$ in Fig.~\ref{fig:scattering}(a) and $\Omega=0.1$ in Fig.~\ref{fig:scattering}(b)].
In general,
the  interaction scenario depends on the whole set of parameters, but
reflection dominates at small velocities ($v\lesssim 0.4$) and relatively large  $\Omega$ and {\it vice versa}   larger velocities (say,  $v\sim 2$) and smaller values of  $\Omega$ favor transmission.

In either of the scenarios presented in Fig.~\ref{fig:scattering}, \SOD induces spin precession [third column of Fig.~\ref{fig:scattering}] whose frequency is given by $\Omega$.  The precession is initiated through the atom transfer between the dark states in the defect region, which changes $S_1$ and perturbs the initial one-soliton solution. The integrability of the system is ``restored'' after the soliton passes the defect, but the soliton is now transformed into a breather characterized by the internal frequency $\Omega$ (for the discussion of the two-soliton solutions of the MS see e.g.~\cite{two-soliton}). Small modification of $\Omega$ also strongly affects the component $S_3(t)$ which acquires nearly constant value after scattering [$\sim 1$ in Fig.~\ref{fig:scattering}(a) and $\sim -1$ in Fig.~\ref{fig:scattering}(b)].

\section{Conclusion}

 Summarizing, we for the first time introduced the system with localized SO coupling. The nontrivial interplay between SO coupling and Zeeman splitting is revealed. These two effects acting together lead to non-integrability of the underlying GP equation and the emergence of a number of nontrivial soliton properties. The central among them is the possibility of formation of stable quasi-scalar soliton complexes, having no analogs in uniform BECs, due to qualitative modification of interaction forces between out-of-phase solitons mediated by SO coupling. The results obtained here may be extended to dissipative exciton-polariton BECs and to optical systems, governed by similar evolution equations.

\acknowledgments
VVK and DAZ acknowledge  support of the FCT (Portugal) grants PEst-OE/FIS/UI0618/2014 and   PTDC/FIS-OPT/1918/2012.

\appendix

\section{On bifurcation of the nonlinear modes}
\label{app:bifurc}

To perform the small amplitude expansion (\ref{eq:small_amp}) we collect all the terms with the same powers of $\epsilon$,  and obtain that in the leading order Eq.~(\ref{w}) is satisfied by $\bw_0$ defined in (\ref{w0}). Proceeding in a way similar to Ref.~\cite{Lakoba} we rewrite the first order equations in the form
\begin{eqnarray}
\label{LF}
\hat{L}W_1(\xi)=F_1(\xi)
\end{eqnarray}
for the vector
\begin{eqnarray}
W_1=\left(\!
\begin{array}{c}
u_1 \\ u_1^* \\ iv_1 \\ -iv_1^*
\end{array}
\! \right).
\end{eqnarray}
The operator $\hat{L}$ is defined by
\begin{eqnarray}
\label{L}
\hat{L}=\left(\!
\begin{array}{cc}
L_u & 0
\\
0 &L_v
\end{array}\right),
\\
L_u= \left(\frac{d^2}{d\xi^2}+4u_0^2(\xi)-1\right)\sigma_0+2u_0^2(\xi)\sigma_1,
\\
L_v= \left(\frac{d^2}{d\xi^2}+2u_0^2(\xi)-1\right)\sigma_0,
\end{eqnarray}
and the right hand side is given by
\begin{eqnarray}
\label{F}
F=\frac{1}{\cosh \xi}\left(\!
\begin{array}{c}
\sin(2\alpha)\cos[Q(\xi)]-\nu_1/2
\\
\sin(2\alpha)\cos[Q(\xi)]-\nu_1/2
\\
\cos(2\alpha)\cos[Q(\xi)]+i\sin[Q(\xi)]
\\
\cos(2\alpha)\cos[Q(\xi)]+i\sin[Q(\xi)]
\end{array}
\!\right).
\end{eqnarray}

Next, defining the inner product between two-column vectors $G_1(\xi)$ and $G_2(\xi)$ by
\begin{eqnarray}
(G_1,G_2)=\int_{-\infty}^{\infty}G_1^\dag(\xi) G_2(\xi)\, d\xi,
\end{eqnarray}
one finds that the kernel $\hat{L}$ is spanned by four orthonormal eigenstates:
\begin{eqnarray*}
	P_1=\frac{1}{2\cosh\xi}\left(\!
	\begin{array}{c}
		1 \\-1 \\0 \\0
	\end{array}
	\!\right),\,\,\,
	P_2=\frac{\sqrt{3}\sinh\xi}{2\cosh^2\xi}\left(\!
	\begin{array}{c}
		1 \\1 \\0 \\0
	\end{array}
	\!\right),
	\\
	P_3=\frac{1}{\sqrt{2}\cosh\xi}\left(\!
	\begin{array}{c}
		0 \\0 \\1 \\0
	\end{array}
	\!\right),\,\,\,
	P_4=\frac{1}{\sqrt{2}\cosh\xi}\left(\!
	\begin{array}{c}
		0 \\0 \\0 \\1
	\end{array}
	\!\right).
\end{eqnarray*}

Existence of a solution of (\ref{LF}) is determined by the Fredholm alternative, i.e. by the requirements $(P_j,F_1)=0$ which must be satisfied for all $j=1,2,3,4$. One readily ensures that $(P_1,F_1)=(P_2,F_1)$ and  $(P_3,F_1)=(P_4,F_1)$, i.e. effectively we have two (generally speaking complex) conditions which are reduced to the conditions (\ref{constratints}). These conditions do not involve $\nu_1$, which  means that $\nu_1=0$ and hence $\mu=\mu_0+{\cal O}(\epsilon^2)$.

\section{On the linear stability analysis}
\label{app:stabil}

 The linear stability analysis was performed by means of substitution of slightly perturbed wavefunctions ($j=1,2$)
 \begin{eqnarray}
 \Psi_{j}=\left[\psi_{jr} (x)+i\psi_{ji} (x)+ (u_{j}+iv_{j})e^{\delta t}\right]e^{-i\mu t}
 \end{eqnarray}
 where the indexes  $r$ and $i$ stand for the real and imaginary parts, $u_{j}$ and $v_{j}$ are the real and imaginary parts of the perturbation, into GPE (\ref{GPE}) and linearizing it with respect to $u_{j}$ and $v_{j}$ which can grow with the complex rate $\delta=\delta_r+i\delta_i$ upon evolution. This linearized eigenvalue problem reads
 \begin{widetext}
 	\begin{eqnarray*}
 		\delta u_1 =-\frac 12 \frac{d^2v_1}{dx^2}+\frac{\kappa^2}{2}v_1+\kappa\frac{du_2}{dx}+\frac 12\frac{d\kappa}{dx}u_2+\frac{\Omega}{2}v_1-\mu v_1
 		-
 		2\psi_{1r}\psi_{1i}u_1-3\psi_{1i}^2v_1-2\psi_{1i}\psi_{2r}u_2-2\psi_{1i}\psi_{2i}v_2-(|\psi_{2}|^2+ \psi_{1r}^2)v_1
 		\\
 		\delta v_1 =\frac 12 \frac{d^2u_1}{dx^2}-\frac{\kappa^2}{2}u_1+\kappa\frac{dv_2}{dx}+ \frac 12 \frac{d\kappa}{dx}v_2-\frac{\Omega}{2}u_1+\mu u_1
 		+ 2\psi_{1r}\psi_{1i}v_1+3\psi_{1r}^2u_1+2\psi_{1r}\psi_{2r}u_2+2\psi_{1r}\psi_{2i}v_2+(|\psi_{1}|^2+\psi_{2i}^2)u_1
 		\\
 		\delta u_2 =-\frac 12 \frac{d^2v_2}{dx^2}+\frac{\kappa^2}{2}v_2+\kappa\frac{du_1}{dx}+\frac 12\frac{d\kappa}{dx}u_1-\frac{\Omega}{2}v_2-\mu v_2
 		-
 		2\psi_{2r}\psi_{2i}u_2-3\psi_{2r}^2v_2-2\psi_{2i}\psi_{1r}u_1-2\psi_{2i}\psi_{1i}v_1-(|\psi_{1}|^2+ \psi_{2r}^2)v_2
 		\\
 		\delta v_2 =\frac 12 \frac{d^2u_2}{dx^2}-\frac{\kappa^2}{2}u_2+\kappa\frac{dv_1}{dx}+ \frac 12 \frac{d\kappa}{dx}v_1+\frac{\Omega}{2}u_2+\mu u_2
 		+ 2\psi_{2r}\psi_{2i}v_2+3\psi_{2r}^2u_2+2\psi_{2r}\psi_{1r}u_1+2\psi_{2r}\psi_{1i}v_1+(|\psi_{1}|^2+\psi_{2i}^2)u_2
 	\end{eqnarray*}
 \end{widetext}
It was solved numerically in order to get the dependence of the perturbation growth rate $\delta$ on the chemical potential $\mu$. The solitons are stable as long as $\delta_r\leq 0$.


\begin{thebibliography}{99}

\bibitem{Lewenstein} M. Lewenstein, A. Sanpera, V. Ahufinger, B Damski, B., A. Sende, and  U. Sen,
Adv. Phys. {\bf 56}, 243 (2007).

\bibitem{hydro} C. Pethick and H. Smith, \textit{Bose-Einstein Condensation
	in Dilute Gases} (Cambridge University Press: Cambridge, 2002); L. P.
	Pitaevskii and S. Stringari, \textit{Bose-Einstein Condensation} (Clarendon
	Press: Oxford and New York, 2003).
	
\bibitem{gravity} L. J. Garay, J. R. Anglin, J. I. Cirac, and P. Zoller, Phys. Rev. Lett.
{\bf  85}, 4643 (2000); C. Barcel\'o, S. Liberati, and M. Visser, Int. J. Mod. Phys. A
{\bf  18}, 3735 (2003).

\bibitem{optics} Y. V. Kartashov, B. A. Malomed, and L. Torner, Rev. Mod. Phys. {\bf 83}, 247 (2011).

\bibitem{Nature} V. Galitski and I. B. Spielman, Nature (London) {\bf 494}, 49 (2013).

\bibitem{Galitski} T. D. Stanescu, B. Anderson, and V. Galitski, Phys. Rev. A
{\bf 78}, 023616 (2008).

\bibitem{Spielman} Y. J. Lin, K. Jimenez-Garcia, and I. B. Spielman, Nature
(London) {\bf 471}, 83 (2011).

\bibitem{Delibard} J. Dalibard, F. Gerbier, G. Juzeli\=unas, and P. \"Ohberg,
Rev. Mod. Phys. {\bf 83}, 1523 (2011).

\bibitem{Kevrekid}  V. Achilleos, D. J. Frantzeskakis, P. G. Kevrekidis, and D. E. Pelinovsky, Phys. Rev. Lett. {\bf 110}, 264101 (2013).

\bibitem{Wu} Y. Xu, Y. Zhang, and B. Wu, Phys. Rev. A {\bf 87}, 013614 (2013).

\bibitem{KKA} Y. V. Kartashov, V. V. Konotop, and F. K. Abdullaev, Phys. Rev. Lett. {\bf 111}, 060402 (2013).

\bibitem{OL}  H. Sakaguchi and B. Li, Phys. Rev. A {\bf 87}, 015602 (2013); Y. Zhang and C. Zhang, Phys. Rev. A {\bf 87},  023611 (2013).


\bibitem{period_exper} K. Jime\'nez-Garc\'ia, L. J. LeBlanc, R. A. Williams, M. C.
Beeler, A. R. Perry, and I. B. Spielman, Phys. Rev. Lett. {\bf 108}, 225303 (2012).

\bibitem{OL_exper} C. Hamner, Yongping Zhang, M. A. Khamehchi, Matthew J. Davis, P. Engels,  	 arXiv:1405.4048 [cond-mat.quant-gas]



\bibitem{Juzel} J. Ruseckas, G. Juzeli\=unas, P. \"Ohberg, and M. Fleischhauer, Phys. Rev. Lett. {\bf 95}, 010404 (2005); G. Juzeli\=unas, J. Ruseckas, M. Lindberg, L. Santos, and P. \"Ohberg,
Phys. Rev. A {\bf 77}, 011802 (2008).

\bibitem{MB} M. V. Berry, Proc. R. Soc. London, Ser. A {\bf 392}, 45 (1984);  F. Wilczek and A. Zee, Phys. Rev. Lett. {\bf 52}, 2111 (1984);  C. A. Mead, Rev. Mod. Phys. {\bf 64}, 51 (1992).


\bibitem{nonlinearity}  Y. Zhang, Li Mao, and C. Zhang, Phys. Rev. Lett. {\bf 108}, 035302 (2012).

\bibitem{Manakov} S. V. Manakov, Zh. Eksp. Teor. Fiz. {\bf 67}, 543 (1974) [Sov. Phys. JETP {\bf 38}, 248 (1974)].









\bibitem{model_optics} D. N. Christodoulides and R. I. Joseph, Opt. Lett. {\bf 13}, 53 (1988);  B. A. Malomed, Phys. Rev. A {\bf 43}, 410 (1991); V. M. Eleonskii, V. G. Korolev, N. E. Kulagin, and L. P. Shil'nikov, Zh. Eksp. Teor. Fiz. {\bf 99}, 1113 (1991) [Sov. Phys. JETP {\bf 72}, 619 (1991)]; M. Haelterman and A. Sheppard, Phys. Rev. E {\bf 49}, 3376 (1994) C. De Angelis and S. Wabnitz, Opt. Commun. {\bf 125}, 186 (1996).

\bibitem{Jianke} J. Yang, \textit{Nonlinear Waves in Integrable and Nonintegrable Systems} (SIAM, Philadelphia, 2010).



\bibitem{two-soliton} R. Radhakrishnan, M. Lakshmanan, and J. Hietarinta, Phys. Rev. E {\bf 56}, 2213 (1997); J. Yang and D. J. Benney, Stud. Appl. Math {\bf 96}, 111 (1996).

\bibitem{Lakoba} T. I. Lakoba and D. J. Kaup, Phys. Rev. E {\bf 56}, 6147 (1997).





















\end{thebibliography}
\end{document}